\global\def\draftcontrol{0}

   \def\versionno{ time -- draft -- 13.08.02   }

\catcode`\@=11

\expandafter\ifx\csname draftcontrol\endcsname\relax\global\def\draftcontrol{0}
\fi

{\count255=\time\divide\count255 by 60
\xdef\hourmin{\number\count255}
\multiply\count255 by-60\advance\count255 by\time
\xdef\hourmin{\hourmin:\ifnum\count255<10 0\fi\the\count255}}
\def\draftdate{\number\month/\number\day/\number\year\ \ \ \hourmin }

\newcommand\makepapertitle{\par
  \begingroup
    \renewcommand\thefootnote{\@fnsymbol\c@footnote}%
    \def\@makefnmark{\rlap{\@textsuperscript{\normalfont\@thefnmark}}}%
    \long\def\@makefntext##1{\parindent 1em\noindent
            \hb@xt@1.8em{%
                \hss\@textsuperscript{\normalfont\@thefnmark}}##1}%
     \newpage
     \global\@topnum\z@   
     \@makepapertitle
     \thispagestyle{empty}\@thanks
  \endgroup
  \setcounter{footnote}{0}%
  \global\let\thanks\relax
  \global\let\makepapertitle\relax
  \global\let\@makepapertitle\relax
  \global\let\@thanks\@empty
  \global\let\@author\@empty
  \global\let\@date\@empty
  \global\let\@title\@empty
  \global\let\title\relax
  \global\let\author\relax
  \global\let\date\relax
  \global\let\and\relax
  \def\version{\let\version\@version\@gobble}
}
\def\@makepapertitle{%
  \newpage
   \ifnum\draftcontrol=1 {}
   \version\versionno
   \vskip 3em%
   \else
   \hfill\hbox to 3cm {\parbox{4cm}{\@pubnum}\hss}%
   \vskip 3em%
   \fi
   \begin{center}%
   \let \footnote \thanks
     {\LARGE {\@title}}%
     \vskip 1.5em%
     {\normalsize
       \lineskip .5em%
       \begin{tabular}[t]{c}%
         \@author
       \end{tabular}\par}%
     \vskip 1em%
     {\@bstract}%
     \end{center}%
     \vskip .5em
     \@date%
   \par
}

\gdef\@pubnum{}
\def\pubnum#1{%
  \gdef\@pubnum{#1}}

\gdef\@bstract{}
\def\Abstract#1{%
  \gdef\@bstract{%
   \parbox{\textwidth-0pc}{%
   \centerline{\bf Abstract}\penalty1000
   \noindent
   \renewcommand\baselinestretch{1.0}
   {#1}}}
}

\def\ps@paper{\let\@mkboth\@gobbletwo%
     \ifnum\draftcontrol=1
        \def\@oddfoot{\hbox to \textwidth{\tiny \versionno \hfil\tiny\draftdate}%
        \hskip -\textwidth \hbox to \textwidth{\hfil\rm\thepage\hfil}}%
     \else\def\@oddfoot{\hbox to \textwidth{\hfil\rm\thepage\hfil}}
     \fi
     \let\@evenfoot\@oddfoot
}

\def\body{\clearpage
          \pagestyle{paper}
        }
\newenvironment{acknowledgments}{%
\vskip 3.25ex
\noindent {\bf Acknowledgments}
}

\def\@version#1{\ifnum\draftcontrol=1
\typeout{}\typeout{#1}\typeout{}
\vskip3mm\centerline{\hbox{\fbox{\normalsize{\tt DRAFT -- #1 -- }
                   {\draftdate}}}}\vskip3mm
\fi}
\let\version\@version
\long\def\eqlabel#1{\ifnum\draftcontrol=1
                    \tag@false  
                    \tag*{(\theequation) \hbox to -0.2cm{\hspace{0cm}\small{#1}\hss}}
                    \refstepcounter{equation} 
                    \edef\@currentlabel{\theequation}
                    \ltx@label{#1}          
                    \else
                    \label{#1}
                    \fi
                    }
\let\st@bibitem\@bibitem
\let\st@lbibitem\@lbibitem
\ifnum\draftcontrol=1
  \def\@bibitem#1{%
    \st@bibitem{#1}\a@@label{#1}\ignorespaces}
  \def\@lbibitem[#1]#2{%
    \st@lbibitem[#1]{#2}\a@@label{#2}\ignorespaces}
  \def\a@@label#1{%
    \gdef\a@lab{\smash{\normalfont\small#1}}
    \ifvmode
      \if@inlabel
        \global\setbox\@labels\hbox{%
          \llap{\a@lab\let\a@lab\relax
                \kern\@totalleftmargin\kern\marginparsep}%
          \box\@labels}%
      \fi
    \fi}
\fi

\documentclass[12pt,letterpaper]{article}

\usepackage{amsmath,amssymb,array,calc,rotating,epsfig,psfrag}
\usepackage[nosort]{cite}

\ifnum\draftcontrol=1
\fi

\renewcommand\baselinestretch{1.25}
\renewcommand\section{\@startsection {section}{1}{\z@}%
                                   {-3.5ex \@plus -1ex \@minus -.2ex}%
                                   {2.3ex \@plus.2ex}%
                                   {\normalfont\large\bfseries}}
\renewcommand\subsection{\@startsection{subsection}{2}{\z@}%
                                   {-3.25ex\@plus -1ex \@minus -.2ex}%
                                   {1.5ex \@plus .2ex}%
                                   {\normalfont\normalsize\bfseries}}
\renewcommand\subsubsection{\@startsection{subsubsection}{3}{\z@}%
                                   {-3.25ex\@plus -1ex \@minus -.2ex}%
                                   {1.5ex \@plus .2ex}%
                                   {\normalfont\normalsize\it}}
\renewcommand\paragraph{\@startsection{paragraph}{4}{\z@}%
                                   {-2.5ex\@plus -1ex \@minus -.2ex}%
                                   {1ex \@plus .2ex}%
                                   {\normalfont\normalsize\it}}
\numberwithin{equation}{section}

\def\ie{{\it i.e.}}
\def\eg{{\it e.g.}}

\def\revise#1       {\raisebox{-0em}{\rule{3pt}{1em}}%
                     \marginpar{\raisebox{.5em}{\vrule width3pt\
                     \vrule width0pt height 0pt depth0.5em
                     \hbox to 0cm{\hspace{0cm}{%
                     \parbox[t]{4em}{\raggedright\footnotesize{#1}}}\hss}}}}

\newcommand\nxt[1]  {\\\fnxt#1}

\def\calc         {{\cal C}}

\def\calm         {{\cal M}}
\def\caln         {{\cal N}}

\def\calw         {{\cal W}}

\def\reals        {{\mathbb R}}
\def\zet          {{\mathbb Z}}

\def\del          {\partial}

\def\sqr#1#2{{\vcenter{\vbox{\hrule height.#2pt  
 \hbox{\vrule width.#2pt height#1pt \kern#1pt
 \vrule width.#2pt}\hrule height.#2pt}}}}

\def\O{\Omega}
\def\a{\alpha}
\def\w{\omega}
\def\r{\rho}

\def\ttheta{\tilde{\theta}}
\def\tphi{\tilde{\phi}}

\def\ZZ {\zet}
\def\SL {{\it SL}}
\def\SU {{\it SU}}
\def\SO {{\it SO}}
\def\U  {{\it U}}

\catcode`\@=12

\begin{document}


\title{On Time-dependent Backgrounds in Supergravity and String Theory}

\pubnum{%
NSF-ITP-02-64 \\
YITP-SB-02-37 \\
hep-th/0207214}
\date{July 2002}

\author{Alex Buchel$^{1}$\footnote{\tt buchel@kitp.ucsb.edu},  
Peter Langfelder$^{1,2}$\footnote{\tt peter.langfelder@sunysb.edu},
and Johannes Walcher$^{1}$\footnote{\tt walcher@kitp.ucsb.edu} \\[0.4cm]
\it $^{1}$Kavli Institute for Theoretical Physics \\
\it University of California \\
\it Santa Barbara, CA 93106, USA \\[0.2cm]
\it $^{2}$C.\ N.\ Yang Institute for Theoretical Physics \\
\it State University of New York \\
\it Stony Brook, NY 11794-3840, USA 
}

\Abstract{
Time-dependent solutions of supergravity and string theory are studied.
The examples are obtained from de Sitter deformation of gauge/gravity
dualities, analytical continuation of static solutions, and ``exactly 
solvable'' worldsheet models. Among other things, it is shown that 
turning on a Hubble parameter in the background of a confining gauge 
theory in four dimensions can restore chiral symmetry. Some of the 
solutions obtained from analytical continuation have the interpretation 
of a universe with a bounce separating a big bang from a big crunch
singularity. In the worldsheet context, it is argued why string
propagation close to a Milne-type cosmological singularity might be 
physically non-singular.
}


\makepapertitle

\body


\version\versionno

\section{Introduction and Outline}

The understanding of space-like singularities is a timely question. More
broadly, one of the important open problems in string theory is the correct
treatment of time-dependent backgrounds. It is generally hoped that there is a
way in which string theory will solve the conceptual and technical problems
associated with time dependence in a full fledged quantum mechanical and
gravitational theory. In this context, the practical point of view, that
we shall adopt, is to neglect the conceptual difficulties and to try to
naively extend the ordinary rules of the game to the time-dependent
setting.

Consider for instance spacetime singularities. In recent years, motivated
and guided by the AdS/CFT correspondence, many new lessons have been learned
about time-like singularities and their resolution in supergravity and string
theory. Firstly, the goal of exhibiting new tests of the duality has led to 
the construction of many new exact supergravity backgrounds. This includes
metrics on non-compact Calabi-Yau spaces with and without fluxes, 
$G_2$-holonomy metrics, etc. Secondly, the correspondence with gauge theory 
has hinted at the correct interpretation and resolution of singularities.
The mechanisms include confinement, chiral symmetry breaking, generation of a
mass gap, etc. These recent advances have complemented the existing knowledge
about singularities from closed strings and D-branes. It is not unreasonable to
hope that these lessons will also teach us something about space-like
singularities.

In this paper, we collect a number of ideas on the treatment of time-dependent 
problems in string theory. One way of introducing time dependence
in the context of the gauge/gravity duality was suggested in \cite{b0203}.
The basic idea of \cite{b0203} (we will review this in detail in section
\ref{desitter}) was to consider a deformation of the ordinary correspondence 
by turning on a Hubble parameter (cosmological constant), so that the 
gauge theory effectively lives on a de Sitter space. In the conformal 
case, this translates simply into a different slicing of the dual AdS space. 
But in non-conformal examples such as $\caln=1$ SYM in four dimensions, 
there is an interesting interplay between the scale of the theory $\Lambda$ 
and the Hubble scale $H$. For instance, for $H\ll\Lambda$, the theory is a 
minor deformation from the flat case. On the other hand, for $H\gg\Lambda$, one
observes the more drastic phenomenon of restoration of chiral symmetry. In 
this respect, the de Sitter deformation of the correspondence is very similar 
to turning on a finite temperature. In \cite{b0203}, these effects were
studied for the de Sitter deformation of the Klebanov-Strassler background,
which realizes $\caln=1$ SYM theory by  D5 and D3 branes on the
conifold. In this paper, in section \ref{mndesitter}, we will consider
the $(1,1)$ little string theory realizing NS5-branes wrapped on $S^2$.
We give a careful discussion of the physics involved and also study chiral
symmetry restoration in the de Sitter background.

In sections \ref{analytical} through \ref{iDbranes}, we give a somewhat more 
general discussion of time-dependent supergravity backgrounds, focusing on 
those that can be obtained from analytical continuation of known static 
solutions. This is motivated in part by the works in refs.\ \cite{g0203} 
and \cite{a0204}. In \cite{g0203}, it was proposed to construct time-dependent 
backgrounds of string theory as spacelike branes. These S-branes are defined 
as the gravitational backreaction of the decay of unstable D-brane systems. 
However, the explicit solutions found in \cite{g0203}, and later in 
\cite{g0204,kmp0204}  (see \cite{behrndt,quevedo1} for earlier work 
on such solutions, recent related work includes \cite{cornalba,sadik,quevedo2}), 
are singular, and the singularities not well understood. As we will see in 
section \ref{iDbranes}, some of these solutions are actually best thought of 
as analytical continuations of ordinary D-branes, or $i$D-branes, for short. 
In this respect, this is an alternative to the strategy used in \cite{a0204}, 
where the analytical continuation of ordinary black holes was considered. A 
different type of analytical continuation is discussed in section \ref{tduality}. 
Explicitly, we will present a method using T-duality at an intermediate step that 
allows the analytical continuation of backgrounds with non-zero (NS-NS) flux.

Finally, in section \ref{milneuni}, we enter the third current circle of ideas
for studying time dependence in string theory, using the exact worldsheet
approach. The most prominent example is probably the so-called Milne universe,
which is a $\ZZ$ orbifold of two-dimensional Minkowski space. The Milne universe
owes its physical interest to its r\^ole in ekpyrotic cosmology \cite{ekpyrosis}, 
and has been studied in a number of recent works \cite{hosteif,bhkn,nekrasov}, 
see also \cite{simon,lms,lawrence,fm0206,hopo} for recent work on related 
null orbifolds. It turns out that the singularity of Milne type also appears 
in an effective description of certain gauged WZW models (see, for instance, 
\cite{eliezer,kr}), which have been intensively studied in the past 
\cite{wbh,efr,dvv,tseytlin,rove,pete,tsva}. Here, using this connection, we
present an observation that can help explain the non-singularity of string 
propagation near Milne-type singularities. The basic physical intuition is 
that near a spacetime singularity, the most relevant physical states are 
the lightest ones. At a Milne-type singularity probed by strings, these states 
are the winding modes of the string. One then observes that the effective 
background metric seen by winding modes in the gauged WZW is less singular 
than that seen by ordinary particles.

We will end the paper in section \ref{summary} with a brief summary.

\section{De Sitter deformation of the gauge/gravity correspondence}
\label{desitter}

The construction of time-dependent backgrounds for string theory is a 
rather complicated problem. As in the static case, one possible starting 
point is a solution of the corresponding supergravity equations of motion. 
Then, given a supergravity solution, one should check whether the truncation 
of the full string theory in this background to the supergravity modes was 
consistent at all. This necessitates small curvatures and small string 
coupling. One then has to face the problems of singularities and the classical
and quantum stability of the background.

In \cite{b0203} it was proposed to use the deformation of the gauge/gravity
correspondence of Maldacena \cite{m9711} (see \cite{a9905} for a review) as a
tool for the construction of a large class of consistent time-dependent string
theory backgrounds\footnote{An interesting alternative approach to generate 
time-dependent backgrounds by exploiting the supergravity duals to confining 
gauge theories was recently discussed in \cite{km0205}.}. The basic idea is to 
replace the flat spacetime background on which the gauge theory is defined by 
a genuinely time-dependent background, while still keeping gravity on the 
brane non-dynamical. This is a deformation of the gauge theory in the 
generalized sense as explained, for instance, in \cite{bt0111}. On the gravity 
side, this deformation requires a change in the asymptotics of the supergravity 
background.  In fact, all of the wrapped brane solutions also belong to the 
class of deformations in which one replaces the flat gauge theory background 
with a general curved space. The major simplification, compared to the direct 
approach outlined above, comes from the fact that the duality allows us to think 
about a certain class of supergravity backgrounds in terms of a dual strongly 
coupled gauge theory. In particular, we can recast questions of singularities, 
stability, consistent truncation to the supergravity modes, etc., in the dual 
gauge theory language where, as we argue below, they appear to be more tractable.

More specifically, consider the simplest gauge/gravity duality realized 
\cite{m9711} by a system of $N$ D3-branes in a flat type IIB string theory 
background. At small 't Hooft coupling $g_s N\ll 1$, the system is best described 
by open strings and realizes $\SU (N)$ $\caln=4$ supersymmetric gauge theory. In 
the limit of strong 't Hooft coupling this gauge theory has a perturbative
description  as type IIB supergravity on $AdS_5\times S^5$, with $N$ units of
RR 5-form flux through the $S^5$. Being an exact correspondence, the duality 
guarantees that {\it any} consistent (physical) deformation on the gauge theory 
side visible in the large $N$ limit should translate into a consistent background
on the supergravity side. Now the simplest way to introduce time dependence is to
deform the correspondence by turning on a non-zero cosmological constant $H\neq 0$
on the gauge theory side. In other words, we wish to replace the flat four-dimensional
Minkowski space $\calm_4$ by four-dimensional de Sitter space $\calm_4^H$, with
the metric
\begin{equation}
(ds_{\calm_4})^2=-dt^2+d\bar{x}^2\;\;\longrightarrow\;\; (ds_{\calm_4^H})^2\equiv 
-dt^2 +\frac{1}{H^2} \cosh^2 Ht\ d\O_3^2\,.
\eqlabel{desd}
\end{equation}
We expect that much like in the usual correspondence, the supergravity dual to
the gauge theory on $\calm_4^H$ will feature $\calm_4^H$ in the asymptotics of the
type IIB supergravity background. The background must therefore be genuinely time
dependent. Provided that the gauge theory on $\calm_4^H$ is a consistent quantum 
field theory, the resulting time-dependent supergravity background will be 
consistent as well. 

Unfortunately, our understanding of interacting quantum field theories (QFT)
in general time-dependent backgrounds is still rather rudimentary. (For
recent developments see \cite{bms0112,ptw0205}.) Thus, it appears that
we are replacing a difficult question of classical consistency of certain
time-dependent supergravity backgrounds with an equally complicated (and
unsolved) problem in the interacting QFT. This is true for example for the
de Sitter deformation of the gauge/gravity correspondence for the $\caln=4$
gauge theory described above. It is not obvious that the $\caln=4$ gauge theory 
exists as a consistent QFT in the de Sitter background. Moreover, the deformed 
supergravity dual appears to be simply a different slice of the original
$AdS_5\times S^5$ background \cite{b0203}. The difficulty for finding a 
simple plausibility argument for the consistency of this gauge theory on 
four-dimensional de Sitter space can be traced back to the fact that this is 
an interacting QFT in which scale invariance has been broken by the 
introduction of a single scale, $H$. Therefore, it is not easy to evaluate 
the importance of the effects on the IR and UV dynamics. As a result, it is 
not clear that $H\neq 0$ is a ``smooth deformation'' of the $H=0$ duality.

Fortunately, one can argue that the situation is not so grim with the deformed
duality for a {\it non-conformal} field theory which already has an intrinsic 
scale $\Lambda$. Let us assume that this gauge theory on $d$-dimensional 
Minkowski spacetime $\calm_d$ is a well-defined QFT at energy scales much larger 
than $\Lambda$. Since the background metric deformation due to $H\ne 0$ in 
\eqref{desd} does not affect short-distance properties of the metric, we expect 
that local field theories on $\calm_d^H$ should have the same ultraviolet dynamics 
as those on $\calm_d$. 

Rather, it seems that the Hubble scale $H$ should be thought of as an infrared 
cutoff on the gauge theory dynamics. This is supported by the non-conformal examples 
of the de Sitter deformed gauge/gravity duality discussed in \cite{b0203}. Indeed, 
de Sitter deformation resolves the infrared singularity in the supergravity dual 
to $(1,1)$ little string theory (LST; for a review of LST, see, \eg, \cite{a9911}) 
and to Klebanov-Tseytlin $\caln=1$ gauge theory \cite{kt0002}. These resolutions 
are similar to the finite-temperature deformations discussed in \cite{ms9710} and 
\cite{b0011}, respectively. Ultimately, this is not a surprise since a vacuum state 
in an accelerating universe has a nonzero Gibbons-Hawking temperature $T_{GH}=H/2\pi$, 
analogous to the Hawking temperature of a black hole. The identification of $H$ 
with an infrared cutoff on the gauge theory is also supported by the conclusions 
reached in related studies of de Sitter gravity in warped compactifications of type
IIB string theory \cite{bhm0112}. 

Furthermore, gauge theories with a finite mass gap $m_{\it gap}\ne 0$ for $H=0$, 
should remain stable when defined on $\calm_d^H$, provided $H/m_{\it gap}$ is 
sufficiently small. The argument is identical to the one given in \cite{ass0201}. 
The original dual supergravity background had a mass gap, and thus a small 
deformation (which can be smoothly turned off) should not produce a tachyonic 
instability.

Altogether, these arguments suggest that non-conformal gauge theories with a mass 
gap, defined on $\calm_d^H$, would have the same IR and UV properties as their $H=0$ 
cousins, provided  $H\ll{\rm min}\{m_{gap},\Lambda\}$. This singles out such theories 
as toy models for studies of consistent time-dependent backgrounds in string 
theory from the perspective of the deformed gauge/gravity duality. 

Basically, there are three approaches for constructing supergravity duals
to confining four-dimensional gauge theories: a mass deformation of $\caln=4$
SYM \cite{ps0003}, the gauge theory on the world-volume of regular and 
fractional D3-branes \cite{ks0007}, and the effective low-energy $d=4$, 
$\caln=1$ SYM theory arising from the compactification of $(1,1)$ LST on 
$S^2$ \cite{mn0008}. These theories have a consistent UV completion and a mass
 gap. Thus, according to the above arguments, we expect that the de Sitter 
deformation of any of these backgrounds would give rise to a consistent 
time-dependent string theory background. By far the simplest of these 
backgrounds is the one discussed by Maldacena and Nu\~nez (MN) 
\cite{mn0008}. In the next section we discuss de Sitter deformations of 
the MN background. We demonstrate that the Hubble parameter $H$ can be 
smoothly turned off. The deformed geometry has small curvatures, small 
string coupling, and asymptotically (the ultraviolet from the perspective 
of the dual gauge theory) approaches the BPS solution of MN. For reasons 
explained above, we expect it to be stable. We also comment on the 
chiral symmetry restoration phase transition for MN gauge theory in de 
Sitter space, relegating a few technical details to the appendix.

\section{De Sitter deformation of the Maldacena-Nu\~nez background}
\label{mndesitter}

We begin with a brief review of the MN construction for the 
supergravity dual to four dimensional $\caln=1$ $\SU(n)$ super
Yang-Mills theory \cite{mn0008}.
 
Recall that (1,1) little string theory \cite{brs9704,a9911}
is realized on the world-volume of NS5-branes in type IIB string theory. 
The theory has sixteen supercharges, and for $n$ NS5-branes 
reduces in the IR to six-dimensional  $\caln=(1,1)$ $\SU (n)$ SYM. 
Consider wrapping NS5-branes on $S^2$. Such a compactification will 
completely break the supersymmetry unless the supersymmetry generators
are appropriately twisted. The correct twisting is obtained 
by embedding the $\SO(2)$ spin connection of $S^2$ into the $\SO(4)\sim 
\SU(2)_R\times \SU(2)_L$ R-symmetry group of the flat LST. 
It is easy to show that by identifying $\U(1)$ connection of the 
$S^2$ with a $U(1)_R\subset \SU(2)_R\subset \SO(4)$ we preserve 
four supercharges, or $\caln=1$ SUSY in the remaining four flat 
directions. Furthermore, the four massless scalars representing 
transverse zero modes of flat NS5-branes upon twisting become spinors 
on the $S^2$. Thus the only massless mode in the IR are the $d=4$ 
gauge fields and the gauginos.

The supergravity dual \cite{mn0008} to this theory has the string frame
metric
\begin{equation}
ds_{st}^2=dx_4^2+n \left(d\rho^2+G^2 d\O^2_2+\frac{1}{ 4}\sum_a(\w_a-A_a)^2\right)\,,
\eqlabel{mn}
\end{equation}
where $\O^2_2$ is a round $S^2$ (parameterized by $(\ttheta,\tphi)$)
which the branes wrap, and $\w_a$ are the $\SU(2)$ left-invariant one 
forms on the $S^3$ (parameterized by $(\theta,\phi,\psi)$) transverse to 
the NS5-branes,
\begin{equation}
\begin{split}
\w_1&=\cos\phi\ d\theta+\sin\phi\sin\theta d\psi\cr
\w_2&=-\sin\phi\ d\theta+\cos\phi\sin\theta d\psi\cr
\w_3&=d\phi\ +\cos\theta d\psi \,.
\end{split}
\eqlabel{1forms}
\end{equation}  
Also in \eqref{mn}, $A_a$ are the $\SU(2)_R$ gauge fields on the $S^2$ 
realizing the twist,
\begin{equation}
\begin{split}
A_1&=a\ d\ttheta\cr
A_2&=-a \sin\ttheta\ d\tphi\cr
A_3&=\cos\ttheta\ d\tphi \,.
\end{split}
\eqlabel{gaugemn}
\end{equation}  
Finally, there is a dilaton $\Phi=\ln g_s$, and an NS-NS 3-form flux
\begin{equation}
H_3=n\left[-\frac{1}{ 4}(\w_1-A_1)\wedge(\w_2-A_2)\wedge(\w_3-A_3)
+\frac{1}{ 4} \sum_a F_a\wedge(\w_a-A_a)\right] \,,
\eqlabel{Hmn}
\end{equation}
where $F_a=dA_a+\epsilon_{abc} A_b\wedge A_c$. Altogether, the background 
is parameterized by three functions $G,a,g_s$ of the radial coordinate 
$\rho \in [0,+\infty)$. The type IIB supergravity equations of motion can 
be solved analytically in this case with the result \cite{cv9707,mn0008}
\begin{equation}
\begin{split}
G^2&=\frac{1}{ 4} \left(4 \r \coth(2 \r)-\frac{4 \r^2}{ \sinh^2 2\r}-1\right)\cr
g_s&=g_0 \left[\frac{2 G}{ \sinh 2\r}\right]^{1/2}\cr
a&=\frac{2 \r}{ \sinh 2 \r} \,,
\end{split}
\eqlabel{mnsol}
\end{equation}
where $g_0$ is a free parameter. The background \eqref{mn}-\eqref{mnsol} is 
nonsingular, has small curvatures in $\a'$ units, and a small string coupling 
provided $n\gg 1$. It correctly reproduces the IR properties of the dual 
gauge theory: confinement, chiral symmetry breaking, and a mass gap in the
spectrum. In the UV (large $\r$), the leading asymptotics of the solution
\eqref{mnsol} are those of the flat LST.  

\paragraph{De Sitter deformation}
  
To study the de Sitter deformation of the MN model, we replace, following 
\cite{b0203},
\begin{equation}
dx_4^2\;\; \longrightarrow\;\; F^2\left[-dt^2+\frac{1}{H^2}\cosh^2 H t\ d\O_3^2\right]
\eqlabel{mndef}
\end{equation}
in the string frame metric \eqref{mn}. The physical reason for introducing 
an extra warp factor $F\equiv F(\r)$ is analogous to the appearance of a 
warp factor $G$ in the holographic dual to LST on $S^2$. In the MN model,
the size of the $S^2$, which is fixed from the LST perspective, becomes 
$\r$-dependent in the supergravity dual. This is simply due to the 
holographic interpretation of the radial coordinate $\r$ as the RG 
scale at which we probe the gauge theory, with large values of $\r$ 
corresponding to the UV (or short distances) in the gauge theory. The growth 
of the corresponding $S^2$ in the supergravity dual simply reflects the 
fact that we ``measure'' little string theory on an $S^2$ of fixed size with 
shorter and shorter reference scales. Also, the decompactification of the 
$S^2$ in the supergravity at $\r\rightarrow\infty$ is necessary to match 
the flat LST UV limit of the MN model. Clearly, the same must be true for 
the $H$-deformation. A fixed scalar curvature ($=12 H^2$) of the gauge 
theory background  $dS_4$ must ``flow'' in the holographic dual. It must
approach zero as $\r\to \infty$, so we expect $F(\r\to\infty)\to \infty$. 

Finally, we assume the same ansatz for the 3-form flux \eqref{Hmn}, 
and $g_s\equiv g_s(\r)$. With this ansatz, the type IIB supergravity equations 
of motion for the deformed MN model are reduced to\footnote{The prime 
denotes derivative with respect to $\r$.} 
\begin{equation}
\begin{split}
0&=\left[\frac{a' F^4}{ g_s^2}\right]'-\frac{a F^4 (a^2-1)}{ g_s^2 G^2} \cr
0&=\left[\frac{(G^2)' F^4}{ g_s^2}\right]'+\frac{F^4}{ 2 g_s^2 G^2}
\left\{ (a^2-1)^2+G^2[(a')^2-4]\right\} \cr
0&=\left[G^2 F^4\left(\frac{1}{ g_s^2}\right)'\right]'
-\frac{F^4}{ 4 g_s^2 G^2}\left\{ (a^2-1)^2+2 G^2[8 G^2+(a')^2]\right\}\cr
0&=\left[\frac{(F^4)' G^2}{g_s^2}\right]'-\frac{12 n H^2 F^2 G^2}
{g_s^2} \,.
\end{split}
\eqlabel{mngs}
\end{equation} 
There is also a first order constraint coming from fixing the 
reparametrization invariance (the choice of $\r$),
\begin{equation}
\begin{split}
0=&F^2 \bigg\{2 G^2\left[8 G^2 \left(g_s'\right)^2+4 g_s^2 \left(G'\right)^2-
4\left(G^2\right)'\left(g_s^2\right)'-g_s^2 \left(a'\right)^2\right]\cr
&+g_s^2\left[\left(a^2-1\right)^2-8 G^2\left(1+2 G^2\right)\right]\bigg\}
+16 g_s G^3 \bigg\{3 g_s G\left(\left(F'\right)^2-n H^2\right)\cr
&+2 (F')^2 g_s^2 \left(\frac{G}{g_s}\right)'  
\bigg\} \,.
\end{split}
\eqlabel{mncons2}
\end{equation}
Notice that \eqref{mnsol} solves \eqref{mngs}, \eqref{mncons2} with 
$F\equiv 1$ and $H=0$. We have not been able to find an analytical 
solution to \eqref{mngs}, \eqref{mncons2} for $H\ne 0$, so we proceed
with the asymptotic analysis and numerical interpolation.  

There are two classes of regular solutions to \eqref{mngs}, \eqref{mncons2},
representing physically distinct deformations of the MN background. With
a radial coordinate $\r\in [0,\infty)$ they are realized as different 
boundary conditions for the warp factors $F$ and $G$. Namely, the two 
possibilities are
\begin{equation}
\begin{split}
{\rm (a):}\qquad F(\r=0)&=0\qquad G(\r=0)\ne 0\cr
{\rm (b):}\qquad F(\r=0)&\ne 0\qquad G(\r=0)=0\,.
\end{split}
\eqlabel{cases}
\end{equation}

As we show below, to obtain a nonsingular solution with the boundary
condition (b) we must turn on nonabelian gauge fields \eqref{gaugemn}
(i.e. $a(\r)$ is nontrivial), as opposed to turning on $A_3$ only.
This is easily interpreted if we recall what happens in the supersymmetric 
solution, $H=0$, $F=1$. There, non-singularity requires turning on a 
non-trivial $a$, which breaks the $\zet_{2N}\subset U(1)_\phi$ corresponding 
to shifts in $\phi$ in \eqref{1forms} down to $\zet_2\subset U(1)_{\phi}$. 
(The first breaking to $\zet_{2N}$ is due to worldsheet instantons.)
The result is in complete agreement with the dual low energy 
effective $\SU(N)$ SYM, where the $U(1)_{\phi}$ is identified with 
the chiral $U(1)_R$ symmetry, broken by the gaugino condensate to a 
$\zet_2\subset\zet_{2N}\subset U(1)_R$ in the IR, \cite{mn0008}.
Thus we identify boundary condition (b) in \eqref{cases} as 
corresponding to $\caln=1$ $\SU(N)$ SYM theory in $dS_4$ in a phase of 
spontaneously broken chiral symmetry. It is this phase of the gauge theory 
that we expect to correspond to a consistent time-dependent string theory
background in the supergravity dual, at least for small $H$.

What is the physics corresponding to the boundary condition (a) in 
\eqref{cases}? From the gauge theory perspective, increasing $H$ should
raise the IR cutoff on the dynamics. Thus, for $H$ large enough we 
would expect restoration of the chiral symmetry, similarly
to finite temperature restoration of the chiral symmetry in the MN
model demonstrated in \cite{bf0103,gtv0108}. Indeed, this is
precisely what is realized by the de Sitter deformed MN model with
boundary conditions (a). We delegate the analysis to Appendix \ref{appa},
and only mention the results here. Unlike the case of boundary
condition (b), here there are regular solutions to \eqref{mngs}
both for nontrivial $a$ and for $a\equiv 0$. We don't have much to say
about solutions with nontrivial $a$. They break chiral symmetry, and since 
they do not exist in the $H\to 0$ limit, they can not be identified with 
excitations of the MN model\footnote{In the original MN model there is a 
large class of regular non-BPS excitations with both finite and infinite 
four-dimensional energy density, and broken chiral symmetry \cite{gtv0108}. 
Here, we are finding that there is also an additional large set of globally 
regular solutions with broken chiral symmetry, once the Hubble parameter 
is turned on, $H\ne 0$ .}. 

The solutions with boundary conditions (a) and $a\equiv 0$ are much 
more interesting. They exist as a globally regular solutions only 
for $H> H_{\rm critical}$, where $H_{\rm critical}$ depends on the 
radius of the $S^2$ on which the NS5-branes are wrapped. These solutions 
thus realize the restoration of the chiral symmetry of the gauge theory 
on the de Sitter background in the supergravity context. Finally, we 
mention that we have found a special solution for $H=H_{\rm critical}$. 
The details can be found in Appendix \ref{appa}.
 
\paragraph{Broken Phase}

We continue with the analysis of boundary conditions (b) in 
\eqref{cases}. For $F(\r=0)\ne 0$, by rescaling the Hubble parameter 
$H$ we can always set $F(\r=0)=1$. There is then a three-parameter 
family $\{H,A,g_0\}$ of regular solutions at $\r\to 0$, with 
asymptotics
\begin{equation}
\begin{split}
F&=1+\frac{n H^2 }{2} \r^2+O(\r^4)\cr
G^2&=\r^2\left(1+\left(-\frac{2}{9}-\frac{2 n H^2}{3}-\frac{A^2}{2}\right)\r^2
+O(\r^4)\right)\cr
g_s&=g_0\left(1-\frac{4+3 A^2}{12}\r^2+O(\r^4)\right)\cr
a&=1+A\r^2+O(\r^4) \,.
\end{split}
\eqlabel{mnH0}
\end{equation}  
These solutions are continuously connected to the supersymmetric,
nonsingular MN solution, for which
\begin{equation}
\{H,A,g_0\}_{MN}=\left\{0,-\frac{2}{3},g_0\right\} \,.
\eqlabel{mnslice}
\end{equation}
Here, $g_0$ is the string coupling at $\r=0$. In addition to this 
supersymmetry preserving solution \eqref{mnslice}, there is an infinite 
discrete sequence of non-BPS excitation in the MN model with finite 
four-dimensional energy density \cite{gtv0108}, $\{H,A,g_0\}=\left\{0,a_k,
g_0\right\},\ k=0,1,\ldots$ where $a_k\in (-2,-2/3]$ and $a_0=-2/3$. 
The generic solution $\{H,A,g_0\}=\left\{0,A,g_0\right\}$ has infinite 
four-dimensional energy density \cite{gtv0108}. 

In general, the boundary conditions \eqref{mnH0} describe de Sitter 
deformations of globally regular (non-BPS) excitation of the MN model. 
To see this, let us look at the $\r\to\infty$ asymptotics of a generic 
solution to \eqref{mngs} with $G(\r\to +\infty)\to +\infty$,
\begin{equation}
\begin{split}
F&=(3 n H^2 \r)^{1/2}+\cdots\cr
G^2&=\r+\cdots\cr
g_s&=g_0\left(\r^{3/4} e^{-\r}+\cdots\right)\cr
a&=\Upsilon \r^{-1/2}\left(1+\cdots\right)
+\calc \r^{1/2} e^{-2\r} \left(1+\cdots\right) \,,
\end{split}
\eqlabel{MNI}
\end{equation}
where $\cdots$ denote corrections which are subdominant as $\r\to \infty$.
We have confirmed by numerical integration that $G(\r\to\infty)\to +\infty$, 
provided we start integrating with the  boundary condition (b) in \eqref{cases}.
In this way, the integration constants $\Upsilon$ and $\calc$ in \eqref{MNI} 
become functions of $H$, $A$ in \eqref{mnH0}.

Now apart from $F$ (which we argued on physical grounds must diverge 
as $\r\to \infty$), all the remaining functions have the same leading 
asymptotics as a generic $\{0,A,g_0\}$ solution \cite{gtv0108}. In fact,
apart from $a$, the leading asymptotics coincide with the BPS solution of 
MN. Moreover, the gauge potential $a$ has the same leading asymptotics 
as in the MN solution provided
\begin{equation}
\Upsilon=0 \,.
\eqlabel{fenergy}
\end{equation}

In \cite{gtv0108}, the condition \eqref{fenergy} was found to be necessary
for the non-BPS excitation to have finite four-dimensional energy density. In 
other word, the discrete set of parameters $a_k$ are just the zeros of 
$\Upsilon(H=0,A)$,
\begin{equation}
\Upsilon(H=0,A=a_k)=0,\qquad k=0,1,\ldots \,.
\eqlabel{fenergyt}
\end{equation}
However, this ``energy'' of the gravitational background was computed in
\cite{gtv0108} using the general definition of \cite{hh9501}, valid for
static backgrounds. In our time-dependent case, it is not clear why we
should impose \eqref{fenergy} based on considerations of energy, which is
not a conserved quantity in de Sitter space. From the perspective of
getting a consistent time-dependent string theory background what we need
is that the de Sitter deformed geometry smoothly reduces to the
supersymmetric MN background as $H\to 0$.  This will be the case for
{\it any}  $A=-\frac{2}{3}+O(n H^2)$. Generically, $\Upsilon(H,A=-\frac{2}
{3}+O(n H^2))=O(n H^2)\ne 0$. Actually, it must in principle also be possible 
to modify $A=-2/3$ perturbatively in $n H^2$ to enforce \eqref{fenergy} for 
$H\ne 0$. We verified this numerically for $n H^2\sim 1$. The important
point is that even without this condition, all our solutions reduce for
$H=0$ to regular excitations of the MN model.

To summarize, the analyticity in $H$ of the deformed equations \eqref{mngs}
and of the boundary conditions \eqref{mnH0} guarantees that for 
$n H^2\ll 1$, our solutions will be regular deformations of the 
non-singular MN solution, and thus should describe consistent (and 
in particular stable) time-dependent backgrounds of string theory.   
The same analyticity argument implies that for $n H^2\ll 1$, the
supergravity approximation for the deformed MN geometry is valid
if the supersymmetric MN background is a valid supergravity background.
The same applies to string loop corrections. They are small in the
deformed geometry, provided they were small to begin with.

An interesting open question is what will happen with de Sitter 
deformation of the {\it non-singular} MN solution for $n H^2\gg 1$? 
Recall that if we deform starting from the non-singular solution,
chiral symmetry remains broken. On the other hand, from the dual
gauge theory perspective, we expect chiral symmetry restoration in
the regime $n H^2\gg 1$. Such chirally symmetric solutions indeed 
exist---they are just solutions with (a) boundary condition in 
\eqref{cases}. In Appendix A we argue that the chirally symmetric 
solution exist only above a certain critical value of the Hubble
parameter $H_{\rm critical}$. We identify this as a signature of 
chiral symmetry restoration in the de Sitter deformed MN model. But:
Can this phase transition be seen from the broken phase as well? 
And if so, what is the critical Hubble parameter in this case? 
We hope to address these issues in a future publication.

\paragraph{Discussion}

We have shown here that the gauge/gravity correspondence of Maldacena 
for UV-consistent gauge theories with a mass gap in the IR is a 
convenient starting point for constructing consistent time-dependent 
string theory backgrounds. Unfortunately, this approach is of little 
practical use in attempts to generate phenomenologically viable 
cosmologies. The reason for this is a technical problem rather than
a conceptual one. One of the simplifications that allowed us to study
the de Sitter deformation of the MN model in some detail was that 
the resulting type IIB equations of motion were ordinary differential 
equations. We will not go into details on this here, but it is 
straightforward to verify that the only time-dependent deformation of 
the Minkowski background which does not introduce time-dependence in 
various warp-factors in the geometry or for the fluxes and the 
dilaton are the gauge theory metric deformations
\begin{equation}
\begin{split}
ds_{R^{d,1}}^2&\to -dt^2 +e^{2 H t}\ ds^2_{R^d}\;,\cr
ds_{R^{d,1}}^2&\to -dt^2 +\frac{1}{H^2}\cosh^2 H t\  ds^2_{S^d} \;.
\end{split}
\eqlabel{posstd}
\end{equation}
The first of these is just a de Sitter deformation in the
accelerating patch, the second is the de Sitter deformation we have
studied here. For generic time-dependent deformations we find, for
instance, that the off-diagonal $\r$-$t$ component of the Ricci tensor
(vanishing in the original background) becomes nonzero. Once time
dependence is required in warp factors (in addition to the radial
dependence), the equations of motion become partial differential
equations. Analyzing the solutions becomes then much more challenging.
Thus for all practical purposes ``time-dependent'' backgrounds from
deformations of the gauge/gravity correspondence are actually
``de Sitter'' backgrounds. But the Universe we live in was not de
Sitter at all times! For one reason, the inflation should have stopped
(or substantially slowed down) over a period of time to allow the large
scale structure formation we observe today. So ultimately, for
constructing time-dependent backgrounds with potential phenomenological
interest, we will have to consider alternative methods.

\section{Time dependence from analytical continuation}
\label{analytical}

In the previous section we discussed time-dependent backgrounds for
string theory from the perspective of the deformation of the Maldacena
gauge/gravity correspondence. The presence of a dual gauge theory
description of the backgrounds, which here is a gauge theory formulated 
in de Sitter space, allowed for a simple physical argument for the
consistency of the constructed supergravity backgrounds. We explained
technical difficulties arising from the attempts of extending this
construction to more generic time-dependent string theory backgrounds,
as would be desirable for phenomenological reasons. More promising
in this regard is the approach recently followed in \cite{a0204},
where it was shown that a double analytical continuation of a 
higher-dimensional Kerr black hole yields a cosmological solution with 
a de Sitter phase ending in a Milne phase at late times.

By itself, analytical continuation of a given string theory background does
not provide insights into its physical properties, such as stability and
consistency of string propagation. Nonetheless, analytical continuation is
a very useful technical tool for (at least) constructing time-dependent 
solutions of the supergravity equations of motion. In the following two 
sections, we will take this approach and ask what other types of ``analytical
continuation'' we can perform, and what are the resulting time-dependent
backgrounds. We concentrate mainly on construction techniques, rather than
attempt to present an exhaustive study of the proposed time-dependent
backgrounds. 

Incidentally, the de Sitter deformations we have studied in the previous 
section can also be thought of as analytical continuation of the gauge theory
background after compactification, in the sequence\footnote{This has also 
been observed by Arkady Tseytlin.}:
\begin{equation}
R^{d,1}\rightarrow S^{d+1} \rightarrow dS_{d+1}\,.
\eqlabel{ourcont}
\end{equation}
But while the authors of \cite{a0204} generated new solutions as analytical
continuations of previously known ones, the chain \eqref{ourcont} requires 
finding a new supergravity solution at the first step, because of the 
compactification.

One reason that the analytical continuations in \cite{a0204} were rather 
straightforward is that the backgrounds discussed there were purely 
gravitational. A generic supergravity solution is supported by fluxes,
and typically, the naively analytically continued metric ends up being 
supported by imaginary fluxes, to which it is hard to assign a physical 
meaning. In the next section we demonstrate how beginning with a certain 
class of backgrounds with fluxes one can perform an analytical continuation 
resulting in backgrounds with only real fluxes. The additional ingredient
that allows achieving this is T-duality or, more generally, mirror 
symmetry.

The idea is to exploit the fact that supersymmetric string compactification
are characterized by (complexified) K\"ahler moduli and complex structure
moduli, which are exchanged under T-duality or mirror symmetry. The only
part that can involve fluxes is the real part of the K\"ahler moduli, which
is the NS-NS two-form potential. Therefore, if the original complex structure
moduli were purely imaginary, the dual background could be purely
gravitational. Of course, to really produce a non-trivial flux this way, 
the original geometry has to involve a certain fibration, with K\"ahler
moduli varying over some base. Also note that in thinking about the dual
space in terms of a purely metrical background, we are neglecting all
stringy (\eg, worldsheet instanton) corrections that are crucial for mirror
symmetry. The basic example for all this is the NS5-brane in type IIB string 
theory, which is T-dual to a certain type IIA background involving $\caln=2$ 
minimal models without fluxes \cite{oova}.

Given such a geometry, $\calw$, that is dual to a background with fluxes, 
$\calm$, we can perform naive analytical continuation resulting in a 
time-dependent, purely gravitational background (neglecting stringy 
corrections), $i\calw$. Finally, dualizing back will result in a new 
time-dependent background with {\it real} fluxes, $\tilde\calm$. 
Schematically, we have,
\begin{equation}
\calm \longrightarrow \calw  \longrightarrow i \calw
\longrightarrow \tilde{\calm} \,. 
\eqlabel{immirror}
\end{equation}

We make this explicit in section \ref{tduality}, where we perform the sequence 
of transformations \eqref{immirror} starting from the exactly soluble string 
theory background describing the throat geometry of the near-extremal flat 
NS5-branes in type IIB string theory.    

Another procedure for constructing time-dependent solutions can be described
as follows. In general, the difficult step in finding supergravity solutions
is to find an ansatz and gauge that can be solved by solving reasonable
(\ie, ordinary) differential equations. Usually, the solutions of these
equations become complex upon analytical continuation. But if instead of
continuing the solutions, one continues the original ansatz, one can
look for real (physical) solutions of the resulting analytically continued 
equations. In terms of the original equations, these are of course imaginary 
solutions, in general. We will follow this procedure in section 
\ref{iDbranes}, where we discuss the analytical continuation of
the usual D-brane solutions. The supergravity equations of motion describing
flat D-branes are ordinary differential equations in the radial coordinate
transverse to the brane. In the continuation, this radial coordinate becomes
the time direction. We will refer to the solutions of these equations as
$i$D-branes. Some of these solutions are special cases of the time-dependent 
solutions recently discussed in \cite{g0204}. We should also point out 
that these solutions can not be S-branes in the sense introduced in 
\cite{g0203}. Certain aspects of obtaining genuine spacelike branes as
the decay process of unstable D-branes will be discussed in a separate
publication \cite{blw}.

\section{Analytical continuation via T-duality}
\label{tduality}

In this section, we illustrate the idea of analytical continuation of
supergravity backgrounds with fluxes following steps \eqref{immirror} with
the example of the throat (near-horizon) geometry of the near extremal
NS5-branes in type IIB string theory.

The relevant NS5-brane geometry was described in \cite{ms9710}. The 
string frame metric is (in units where $\alpha'=1$),
\begin{equation}
ds_{st}^2=dx_5^2+k \bigl(-\tanh^2\rho\,dt^2+d\rho^2\bigr)+k\,d\O_3^2 \;,
\eqlabel{ns5metric}
\end{equation}
where we take the $S^3$ metric
\begin{equation}
d\O_3^2=d\theta^2+\cos^2\theta\,d\psi^2+\sin^2\theta\, d\lambda^2 \,.
\eqlabel{s3param}
\end{equation}
There is also NS-NS two-form potential $B_2$ and dilaton $\Phi$
\begin{equation}
\begin{split}
B_2&=k \sin^2\theta\, d\lambda\wedge d\psi\cr
e^{2\Phi}&=\frac{k}{\mu \cosh^2\rho} \,.
\end{split}
\eqlabel{rerstns5}
\end{equation}
In above equations, $k$ is the number of NS5-branes, and $\mu$ is related 
to the energy density above the extremality. The radial coordinate 
$\rho\in [0,\infty)$. String propagation in this geometry corresponds to 
an ``exact conformal field theory''
\begin{equation}
\calc_1\times \calc_2\times\calc_3\equiv 
 [\SL(2,\reals)/\U(1)]_k\times \SU(2)_k\times \reals^5 \,,
\eqlabel{ns5cft}
\end{equation} 
where  the coset $\calc_1$
\begin{equation}
[\SL(2,\reals)/\U(1)]_k
\eqlabel{sigarwit}
\end{equation}
parameterized by $(\rho,t)$ in \eqref{ns5metric}, is the 2-d black hole
of \cite{wbh}; the 3-sphere $(\theta,\psi,\lambda)$ along with the 
$B_2$-field is described by the $\calc_2\equiv\SU(2)_k$ WZW model;
and $\reals^5$ is a free CFT describing noncompact directions 
along the NS5-branes. The central charges of the three conformal theories
in \eqref{ns5cft} are
\begin{equation}
\left[\frac{3(k+2)}{k}-1+2\cdot \frac{1}{2}\right]\times \left[
\frac{3(k-2)}{k}+3\cdot\frac{1}{2} \right]\times \left[5\cdot \frac{3}{2}
\right]\ =\ 15 \,,
\eqlabel{ccharge}
\end{equation}
respectively. The above background is our starting point, $\calm$, in
\eqref{immirror}.

Notice that we cannot perform any analytical continuation of $\calm$
directly. For instance, consider the naive analytical continuation $\rho\to
i\rho$. This is actually a continuation of the coset CFT $\calc_1$ to a
negative level $k\to -k$. As a result, the total central charge of the
background would no longer be $15$, unless we continue the WZW CFT to
negative level as well. However, the latter continuation must be accompanied
by an analytical continuation of the 3-sphere $S^3\to H_3$,
\begin{equation}
(\theta,\psi,\lambda)\longrightarrow (i\theta,i\psi,\lambda) \,,
\eqlabel{s3h3}
\end{equation}
in order to have a single time direction. But now, from \eqref{rerstns5},
the two-form potential becomes purely imaginary. 

To find a physical analytical continuation of this background, following 
\eqref{immirror}, we have to a find a mirror, purely geometric background, 
$\calw$. In our case the appropriate mirror symmetry is that of the WZW 
CFT
\begin{equation}
\calc_2\longrightarrow \calw_2 \equiv \Bigl[\U(1)\times 
\frac{\SU(2)_{k}}{\U(1)}\Bigr]\Big/ \zet_k \,.
\eqlabel{mirror2}
\end{equation}
At the level of the target space metric, the cover of $\calw_2$ 
corresponds to
\begin{equation}
\begin{split}
ds_{\calw_2}^2&= k\,d\psi^2+ k(d\theta^2+\cot^2\theta\,d\beta^2)\cr
B_2^{(\calw_2)}&=0\cr
\Phi_{\calw_2}&=-\ln\sin\theta+{\rm const}\,,
\end{split}
\end{equation}
where $\psi$ is the $\U(1)$ coordinate, and $(\theta,\beta)$ 
parameterize the $\caln=2$ minimal model in \eqref{mirror2}. The 
$\zet_k$ orbifold generator acts in the usual way. At this level, 
the mirror symmetry \eqref{mirror2} is nothing but T-duality along 
the Killing vector field
\begin{equation}
v_K=\frac{1}{k} \Bigl(\frac{\del}{\del\beta}+\frac{\del}{\del\psi}\Bigr)\,,
\eqlabel{vk}
\end{equation}
where the dual variable is what is called $\lambda$ in \eqref{s3param}.

Now we can implement the third step in \eqref{immirror} by taking
\begin{align}
\calw_2\longrightarrow i\calw_2\qquad
&:\qquad (\psi,\theta,\beta)\longrightarrow (\psi,i\theta,i\beta) 
\eqlabel{cont2}
\intertext{along with} 
\calc_1\longrightarrow i\calc_1\qquad
&:\qquad (\rho,t)\to (\rho,it) \,.
\eqlabel{cont1}
\end{align}

Notice that the transformations \eqref{cont2} and \eqref{cont1}
do not change the central charge, thus $[i\calc_1]\times [i\calw_2]\times 
\calc_3$ is still a critical superstring background. Furthermore,
the background for $i\calw_2$ becomes
\begin{equation}
\begin{split}
ds_{\calw_2}^2&\;\to\; ds_{i\calw_2}^2\equiv k\,d\psi^2 +
k(-d\theta^2+\coth^2\theta\, d\beta^2)\cr
\Phi_{\calw_2}&\;\to\; \Phi_{i\calw_2}\equiv -\ln\sinh\theta+{\rm const}\,,
\end{split}
\eqlabel{metw2}
\end{equation}
which is nothing but ($\U(1)$ times) the T-dual (along $\beta$ direction) of 
an $\SL(2,\reals)/\U(1)$ coset at negative level $-k$. Such a coset was 
recently studied \cite{kr} in the context of strings probing a cosmological 
singularity. We will comment on the cosmological singularity of this coset 
and the relevance of its T-dual \eqref{metw2} in some detail in section
\ref{milneuni} of this paper.

The final step of \eqref{immirror} is just a T-duality of $[i\calc_1]\times 
[i\calw_2]\times  \calc_3$ along the isometry \eqref{vk}. The resulting 
geometry is 
\begin{equation}
\begin{split}
ds_{\tilde\calm}^2&=dx_5^2+k\left(\tanh^2\rho\,dt^2+d\rho^2\right)
-k\,d\theta^2 +\frac{k}{1+\coth^2\theta}\Bigl\{\coth^2\theta\,d\psi^2
+ d\lambda^2\Bigr\}\cr
B_2^{\tilde\calm} = B_2^{(i\calw_2)}&=
\frac{k}{1+\coth^2\theta}\;d\lambda\wedge d\psi\cr
\Phi_{\tilde\calm} = \Phi_{i\calw_2}&=-\ln( \cosh\rho\cosh\theta) -
{\frac 12}\ln(1+\tanh^2\theta)+{\rm const} \,.
\end{split}
\eqlabel{finalstep}
\end{equation}

The background in \eqref{finalstep} has everywhere small curvature in 
string units for $k\gg 1$ and has small string coupling for large negative 
constant in the dilaton expression. Moreover, it corresponds to an exactly 
soluble CFT. We do not have a simple argument settling the question of
stability of \eqref{finalstep}. Note that the starting point $\calm$,
eq.\ \eqref{ns5metric}, is {\it unstable}, as shown in \cite{lstinst}. 
However, it is not clear whether this instability has anything to do with 
instabilities of $\tilde\calm$. For example, one could literally repeat 
the steps of \eqref{immirror} starting from $\caln=2$ supersymmetric 
background corresponding to NS5-branes wrapped on $S^2$ \cite{gkmw0106,
bcz0106,kh0203}. Finally, we note that the metric written in 
\eqref{finalstep} has an apparent singularity at $\theta=0$, as the 
$S^1$ parameterized by $\lambda$ shrinks to zero size. This is a Milne-type 
coordinate singularity, some thoughts about which are presented in section 
\ref{milneuni} below.

\section{$i$D-branes}
\label{iDbranes}

We now turn to the alternate method for constructing analytically continued
solutions from known ones, outlined at the end of section \ref{analytical}.
Specifically, we will apply this method starting from the usual D-brane
solutions \cite{host}. Recall that flat D-branes (or NS5-branes) can be 
obtained as solutions of dilaton-Einstein-Maxwell gravity with the 
Einstein-frame action 
\begin{equation}
S=\int d^{10}x \sqrt{-g} \Bigl(R-\frac{1}{2} (\del\Phi)^2
-\frac{1}{2 q!}e^{a \Phi} F^2_{[q]}\Bigr)\,.
\eqlabel{dem}
\end{equation}
Here, $F_{[q]}$ is the $q$-form field strength sourced by the brane,
and $a$ is the dilaton coupling. For the NS5-brane, $q=3$ and $a=-1$, and 
for a  D$_{8-q}$-brane, $a=(5-q)/2$. The equations of motions derived from 
\eqref{dem} are
\begin{equation}
\begin{split}
R_{\mu\nu}-\frac{1}{2}\del_{\mu}\Phi\del_{\nu}\Phi-\frac{e^{a \Phi}}
{2(q-1)!}\left[F_{\mu\alpha_2\cdots \alpha_q}F_{\nu}\ ^{\alpha_2
\cdots \alpha_q}-\frac{q-1}{8 q} F^2_{[q]} g_{\mu\nu}\right]&=0\cr
\del_{\mu}\left(\sqrt{-g}e^{a \Phi} F^{\mu\alpha_2\cdots \alpha_q}\right)&=0\cr
\frac{1}{\sqrt{-g}}\del_{\mu}\left(\sqrt{-g}\del^{\mu}\Phi\right)
-\frac{a}{2 q!}e^{a \Phi}F^2_{[q]}&=0 \,.
\end{split}
\eqlabel{eomsb}
\end{equation} 

The metric ansatz for a flat D$_{8-q}$-brane is
\begin{equation}
ds_{E}^2=e^{2A }d\rho^2+e^{2B}\left(-dt^2+dx_{8-q}^2\right)+e^{2C} d\O_{q}^2\,,
\eqlabel{metricdbranes}
\end{equation}
where $d\O_{q}^2$ is a metric on a round $S^q$ of unit size, and the warp 
factors $A$, $B$, and $C$ depend only on the radial coordinate $\rho$. In 
addition, the field strength is taken to be
\begin{equation}
F_{[q]}=b\;{\rm vol}(S^q) \,,
\eqlabel{flux}
\end{equation}
where $b$ is a constant related to the total charge of the brane. Finally,
the dilaton $\Phi$ depends on $\rho$ only. For the above ansatz the supergravity
equations of motion \eqref{eomsb} reduce to a coupled system of second order
ordinary differential equations in the variable $\rho$ ($p\equiv 9-q$)
\begin{equation}
\begin{split}
p\left(B''+(B')^2-B' A'\right)+q\left(C''+(C')^2-C' A'\right)+\frac{1}{2}
\left(\Phi'\right)^2-\frac{(q-1)b^2}{16}e^{a \Phi+2 A-2 q C}&=0\cr
B''-B'A'+p\left(B'\right)^2+q B' C'-
\frac{(q-1)b^2}{16}e^{a \Phi+2 A-2 q C}&=0\cr
C''-C' A'+q \left(C'\right)^2+p B' C'-e^{2 A-2 C}(q-1)+
\frac{p b^2}{16}e^{a \Phi+2 A-2 q C}&=0\cr
\left(e^{p B+q C-A} \Phi'\right)'-\frac{1}{2}a b^2 e^{a \Phi+A-q C+p B}&=0 \,.
\end{split}
\eqlabel{dbraneeq}
\end{equation}
The equations \eqref{dbraneeq} have the familiar D-brane solution \cite{host}
\begin{equation}
\begin{split}
\Phi&=\frac{q-5}{4} \ln H\cr
A&=\frac{p}{16}\ln H\cr
B&=\frac{1-q}{16}\ln H\cr
C&=\ln \rho+\frac{p}{16}\ln H\,,
\end{split}
\eqlabel{dbsolve}
\end{equation}
with
\begin{equation}
H=1+\frac{b}{(q-1)\rho^{q-1}} \,.
\eqlabel{hsolve}
\end{equation}

Can we analytically continue these solutions to time-dependent ones? Let
us consider taking
\begin{equation}
(t,\rho)\longrightarrow (i \rho,it) \,.
\eqlabel{dbim}
\end{equation}
Under \eqref{dbim} the solution \eqref{dbsolve} becomes unphysical. 
Indeed, $e^{2C}\to - e^{2C}$, and so in order to have a single time
direction, we have to analytically continue the $q$-sphere transverse to the
D-branes to the $q$-dimensional hyperbolic space, $S^q\to H_q$. If $q$ is
odd  as in type IIB string theory, ${\rm vol(S^q)}\to i {\rm vol}(H_q)$. In
addition, reality of the harmonic function $H$ requires real $b$. Thus the 
$q$-form flux \eqref{flux} becomes purely imaginary. On the other hand, if 
$q$ is even as in type IIA string theory, ${\rm vol(S^q)}\to {\rm vol}(H_q)$. 
But now the reality of $H$ requires $b\to i b $---so again we end up with an 
imaginary flux \eqref{flux}. 

A very simple way to circumvent the above obstacle is to analytically continue  
the supergravity {\it equations} \eqref{dbraneeq}, rather than their
{\it solutions} \eqref{dbsolve}! Clearly, the analytically continued equations 
are real and their solutions describe time-dependent backgrounds of the form
\begin{equation}
ds_{E}^2=-e^{2A }dt^2+e^{2B}\left(d\rho^2+dx_{8-q}^2\right)+e^{2C} d\O_{q}^2\,,
\eqlabel{metricdbranest}
\end{equation}
with fluxes given by \eqref{flux}. Now all the warp factors and the dilaton
depend on time. The analytically continued equations may have real (physical!)
solution which were missed by the analytical continuation of the original
solutions. The D-brane example discussed in this section is precisely of this
type. In the gauge
\begin{equation}
-A+p B +q C=0\,,
\eqlabel{ggauge}
\end{equation}
the analytically continued equations \eqref{dbraneeq} become just the
equations studied in \cite{g0204} in the context of constructing supergravity
solutions corresponding to S-branes. More precisely, the solutions
found in \cite{g0204} fall in different classes depending on the signature
of the curvature of the transverse space. The solutions \eqref{metricdbranest}
are those that have $\sigma=+1$ in \cite{g0204} (spherical transverse space).

We mention that these $i$D-branes are rather different from the spacelike 
branes proposed in \cite{g0203}. Indeed, the solutions
\eqref{metricdbranest} have an $\SO(q+1)$ ``R-symmetry'', instead of
$\SO(q,1)$ spontaneously broken down to $\SO(q)$, as suggested by the
original arguments in \cite{g0203}. Moreover, the fluxes sourced by
these branes do not match the S-brane requirements. As explained in
\cite{g0203}, even and odd codimension S-branes should have physically very
different fluxes. The former should source a flux spreading only {\it on}
their lightcone, while the flux produced by the latter should spread {\it
inside} the entire lightcone of the branes. Here, the flux is always
distributed over the entire sphere. Finally, it is easy to see that the
ansatz \eqref{metricdbranest} does not admit any asymptotically flat
solution, and so is unlikely to correspond to the dynamical decay of an
unstable D-brane system in string theory, which should produce just the
closed string vacuum. 

The physical meaning of the $i$D-branes (apart from the fact that they
are supergravity backgrounds) is somewhat unclear. One of their properties
is that they have a Big Crunch singularity in the future and a Big Bang 
singularity in the past \cite{g0204}. To close this section, we briefly 
discuss a less trivial example of our solution-generating technique. 
Namely, we present solutions corresponding to wrapped versions of the 
above $i$D-branes. The motivation to construct these solutions is to check 
whether or not the wrapping can resolve the cosmological singularity of the 
flat branes. Explicitly, we will consider $i$NS5-branes wrapped on $S^2$
obtained from analytical continuation of the MN background \cite{mn0008}.
The result is a time-dependent supergravity background with five non-compact
directions\footnote{Applying this analytical continuation to NS5-branes 
wrapped on $S^3$ constructed in \cite{mn} gives rise to a four-dimensional 
time-dependent superstring background.}.

\paragraph{$i$NS5 wrapped on $S^2$}

The supergravity equations of motion describing the MN background are given 
in \eqref{mngs} with $F\equiv 1$ and $H=0$. The analytically continued equations
$\rho\to it$ and $t\to i\rho$ are
\begin{equation}
\begin{split}
0&=\left[\frac{a'}{ g_s^2}\right]'+\frac{a (a^2-1)}{ g_s^2 G^2} \cr
0&=\left[\frac{(G^2)'}{ g_s^2}\right]'-\frac{ (a^2-1)^2-G^2[(a')^2+4]}{ 2 g_s^2 G^2} \cr
0&=\left[G^2 \left(\frac{1}{ g_s^2}\right)'\right]'
+\frac{ (a^2-1)^2+2 G^2[8 G^2-(a')^2]}{ 4 g_s^2 G^2}\cr
0&=2 G^2\left[8 G^2 \left(g_s'\right)^2+4 g_s^2 \left(G'\right)^2-
4\left(G^2\right)'\left(g_s^2\right)'-g_s^2 \left(a'\right)^2\right]\cr
&-g_s^2\left[\left(a^2-1\right)^2-8 G^2\left(1+2 G^2\right)\right] \,.
\end{split}
\eqlabel{cons2s2}
\end{equation}
These equations describe the time-dependent background
 \begin{equation}
ds_{st}^2=dx_4^2+n \left(-dt^2+G^2 d\O^2_2+\frac{1}{ 4}\sum_a(\w_a-A_a)^2\right)\,,
\eqlabel{msms2}
\end{equation}
with a 3-form flux given by \eqref{Hmn}, and a time-dependent string coupling 
$g_s$. 

For simplicity, we only consider solutions invariant under $t\leftrightarrow 
-t$. Nonsingular solutions to \eqref{cons2s2} are characterized by two 
integration constants $g_0\equiv g_s|_{t=0}$ and $G_0\equiv G|_{t=0}$.
The asymptotics are
\begin{equation}
\begin{split}
G&=G_0+\frac{4 G_0^2+1}{2 G_0} t^2+O(t^4)\cr
g_s&=g_0\left(1+\frac{4 G_0^2+1}{2 G_0} t^2+O(t^4)\right)\cr
a&=a_0-\frac{a_0(a_0^2-1)}{G_0^2}t^2++O(t^4) \,,
\end{split}
\eqlabel{seriessol}
\end{equation}          
where
\begin{equation}
a_0=\sqrt{1\pm 2\sqrt{4 G_0^4+2 G_0^2}} \,.
\eqlabel{aaadef}
\end{equation}

Notice that $G$ has positive second derivative at $t=0$, thus the boundary
condition \eqref{seriessol} describes a cosmological solution of \eqref{cons2s2}
with a bounce at $t=0$. What happens to our solutions at late/early times?
We verified numerically for both branches of \eqref{aaadef} that the bounce 
always ends up  in a Big Crunch and starts at a Big Bang singularity.  
Physically, it is not surprising that the bounce solutions is singular.
We would expect that a nonsingular solution asymptotically would correspond to 
decompactification of $S^2$, \ie, would describe  the flat $i$NS5  brane of 
\eqref{metricdbranest}.
It is straightforward to verify that there is no solution of  \eqref{metricdbranest}
with $B=C+{\rm const}$, which follows from the metric ansatz \eqref{msms2}. 
On the other hand, it can be checked that \eqref{msms2} is the most general 
metric ansatz invariant under time reversal.

\section{Strings in cosmological WZW models and Milne universe}
\label{milneuni}

In this section, we analyze a worldsheet description of strings at a
cosmological singularity. The model we study is related to the well-known
Witten's black hole background \cite{wbh} by analytic continuation. Hence,
it is equivalent to a certain gauged WZW model, based on
$\SL(2,\reals)/\U(1)$, analytically continued to negative level. The
associated sigma-model is a two-dimensional cosmology, with a singularity
of Milne type, \ie, it is locally of the form
\begin{equation}
ds^2 = -dt^2 + \lambda^2 t^2 dr^2 \,,
\eqlabel{milne}
\end{equation}
where $r$ is periodic with period $2\pi$.

The Milne singularity and similar background have been discussed recently in
a number of works \cite{hosteif,bhkn,CornalbaCosta,nekrasov,simon,lms,
lawrence,fm0206,hopo, eliezer,kr}. Since the singularities are orbifolds, 
one is tempted to expect that a natural extension of classical methods should 
suffice. However, it has not been possible so far to fully understand the 
singularity using all available methods from ordinary Euclidean orbifolds. 
For example, the deformation of the theory by a tachyon v.e.v.\ does not 
smooth the metric \cite{nekrasov}. The existence of closed timelike curves 
poses genuine problems for physics \cite{eliezer,kr}. Moreover, the null 
orbifolds of \cite{lms} (which do not suffer from closed timelike curves) 
are unstable to timelike orbifolds \cite{lms,lawrence,hopo}.

The purpose of this section of our paper is to point out that there is a 
surprising way in which string propagation near the Milne singularity,
obtained as a limit of an ``exactly solvable'' WZW model, might be less 
singular than at first expected. More precisely, we observe that the {\it 
winding modes} of the string see a non-singular background metric. Since 
winding modes are the lightest near the singularity, it is physically 
reasonable to use those as a probe of the geometry, and of the presence of 
a singularity. We hasten to emphasize, however, that the following observations 
can not be viewed as compelling evidence, let alone a proof, that 
strings fully resolve the singularity. For example, we do not address 
the problems associated with closed timelike curves. Moreover, the string
coupling in the winding mode picture appears to diverge, invalidating 
string perturbation theory. Nonetheless, we pursue the question what is the
cosmological interpretation of the gauged WZW model, and, assuming that the
latter is well-defined, why is the cosmology.

\begin{figure}
\begin{center}
\psfrag{u}{$u=0$}
\psfrag{v}{$v=0$}
\psfrag{uv}{$uv=1$}
\epsfig{file=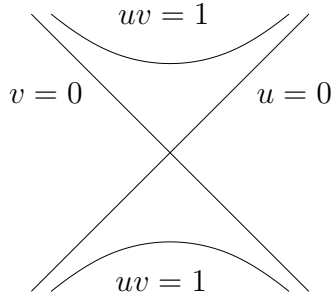,width=4cm}
\caption{Penrose diagram for the metric \eqref{tdbh}. Depending
on the sign of $k$, this represents a black hole or a cosmological
background.}
\label{penrose}
\end{center} 
\end{figure}

Let us begin by recalling the two-dimensional black hole metric and
dilaton \cite{wbh}
\begin{equation}
ds^2 = -k\, \frac{du\, dv}{1-uv}\,, \qquad\qquad
\Phi = -{\textstyle\frac14}\ln (1-uv)^2 \,,
\eqlabel{tdbh}
\end{equation}
where $k>0$. The Penrose diagram for this metric is depicted in fig.\ 
\ref{penrose}. A change of coordinates brings the metric into the
form
\begin{equation}
ds^2 = k (d\rho^2 - \tanh^2\rho\, d\tau^2)\,, \qquad\qquad
\Phi =  -\textstyle{\frac12}\ln \cosh^2\rho \,,
\eqlabel{afbh}
\end{equation}
which covers the asymptotically flat regions of spacetime. An analytical
continuation $t\to i\theta$ yields the popular semi-infinite cigar metric,
but this is not the analytical continuation that we are interested in here.
Instead, we note that, as was exploited for instance in \cite{tsva}, the 
metric \eqref{afbh} with {\it negative} $k$ has a cosmological 
interpretation: The scale factor of space goes to zero in a finite 
proper time. Of  course, there is no real singularity there unless one
compactifies space, as we will do in a moment. The maximally extended 
spacetime then is just fig.\ \ref{penrose} rotated by $90^\circ$.

The important point of ref.\ \cite{wbh} was that the metric \eqref{afbh}
and its analytical continuation to the semi-infinite cigar can be obtained
as the leading order effective spacetime metric associated with a certain
gauged WZW model \cite{efr}. More precisely, the group $\SL(2,\reals)$ has
two non-degenerate subgroups, a compact $\U(1)_-$ of negative signature,
and a non-compact $\U(1)_+$ of positive signature (we endow the $\SL(2,\reals)$ 
group manifold with the metric of signature $(+,+,-)$). Starting from
the WZW model based on $\SL(2,\reals)$ at level $k>0$, one can construct new
models by gauging the various subgroups. Gauging $\U(1)_-$ can obviously 
only give the Euclidean black hole, while gauging $\U(1)_+$ gives 
\eqref{afbh}.

Furthermore, for each of these subgroups, there is the choice of axial
or vector gauging. As discussed in detail in \cite{dvv,rove,pete}, the
two ways of gauging are expected to result in isomorphic conformal field 
theories up to charge conjugation of the left-moving sector of the string. 
But the effective metrics associated with axial and vector gauging are 
different, rather as in ordinary T-duality or mirror symmetry. For 
instance, the metric dual to \eqref{afbh} is given by
\begin{equation}
ds^2 = k (d\rho^2 - \coth^2 \rho \,d\tilde\tau^2) \,,\qquad\qquad
\Phi =-\textstyle{\frac 12} \ln\sinh^2 \rho\,,
\eqlabel{dualbh}
\end{equation}
which can be viewed as local T-duality on the $\tau$ isometry.

As explained above, our interest is to view \eqref{afbh} or \eqref{dualbh}
as a cosmological background, which formally amounts to taking $k$ negative
and writing
\begin{align}
ds^2 &= |k| (-dt^2 + \tanh^2 t \,dr^2) \,, \qquad\qquad
\Phi = -\textstyle{\frac 12}\ln\cosh^2 t \,, \eqlabel{tdco} \\
ds^2 &= |k| (-dt^2 + \coth^2 t \,dr^2) \,, \qquad\qquad
\Phi = -\textstyle{\frac 12}\ln\sinh^2 t\,, \eqlabel{dualtdco}
\end{align}
respectively. Compactifying the $r$ direction and taking the limit 
$t\rightarrow 0$ obviously results in the Milne universe \eqref{milne}
and its conical singularity \cite{kr}, or its T-dual version with a 
curvature singularity, \ie,
\begin{equation}
ds^2 = -dt^2 + \lambda^{-2} t^{-2} dr^2 \,,\qquad\qquad
\Phi =- \ln |t| \,.
\eqlabel{dualmilne}
\end{equation}

It is a reasonable question to ask whether any stringy mechanism could
resolve or explain the apparent spacelike singularities in either of 
\eqref{milne}, \eqref{tdco}, \eqref{dualtdco}, or \eqref{dualmilne}. 
We will here present an observation concerning the possibility that
$\alpha'$ corrections respectively the presence of winding modes can
regulate the singularity.

As reviewed above, the metrics \eqref{afbh}, \eqref{dualbh} appear in the
$\sigma$-model that to first order governs the effective dynamics of
strings described by the gauged WZW model. This was derived in \cite{wbh}
by simply integrating out the gauge field, which is exact in the $k\rightarrow
\infty$ limit. In \cite{dvv}, a careful comparison of the zero-mode 
energy of the string with the target-space Laplacian revealed that the 
exact metrics should in fact be
\begin{align}
ds^2 &= k \bigl(d\rho^2 - \frac{\tanh^2\rho}{1-\frac 
2k\tanh^2\rho}\,d\tau^2\bigr) \,, \!\!\qquad\qquad
\Phi =-\textstyle{\frac 12} \ln \cosh^2\rho \,
(1-{\textstyle\frac 2k}\tanh^2\rho)^{1/2}\,,
\eqlabel{exactbh} \\
ds^2 &= k \bigl(d\rho^2 - \frac{\coth^2\rho}{1-\frac 
2k\coth^2\rho}\,d\tau^2\bigr) \,, \!\qquad\qquad
\Phi =-\textstyle{\frac 12} \ln \sinh^2\rho \,
(1 -{\textstyle\frac 2k}\coth^2\rho)^{1/2}\,.
\eqlabel{dualexactbh}
\end{align}
It was shown in \cite{tseytlin} that the first few terms in a $1/k$
expansion of these expressions solve the known low-loop beta-functions of 
the $\sigma$-model. Since $k$ has dimensions of (length)$^2$, these
corrections should therefore be viewed as $\alpha'$ corrections of the
$\sigma$-model metric. Writing the metrics \eqref{exactbh},
\eqref{dualexactbh} might then seem like a resummation of all $\alpha'$
corrections, which is particularly touchy for \eqref{dualbh}, where the
curvature diverges. However, in the absence of a general rule for
extracting a metric from an ``exactly solvable'', algebraically
defined, conformal field theory model such as the gauged WZW model,
the method used in \cite{dvv} seems a reasonable way to proceed.

We now make the same naive analytical continuation as before, and
compactify $r$. This yields the $\alpha'$ corrected versions of 
\eqref{tdco} and \eqref{dualtdco},
\begin{align}
ds^2 &= k (-dt^2 + \frac{\tanh^2t}{1+\frac 
2k\tanh^2t}\,dr^2\bigr) \,, \!\!\!\!\qquad\qquad
\Phi =-\textstyle{\frac 12} \ln \cosh^2t \,
(1 +{\textstyle\frac 2k}\tanh^2t)^{1/2}\,,
\eqlabel{exacttdco} \\
ds^2 &= k \bigl(-dt^2 + \frac{\coth^2t}{1+\frac 
2k\coth^2t}\,dr^2\bigr) \,, \!\!\!\!\qquad\qquad
\Phi =-\textstyle{\frac 12} \ln \sinh^2t \,
(1 +{\textstyle\frac 2k}\coth^2t)^{1/2} \,.
\eqlabel{dualexacttdco}
\end{align}

The interesting aspect of these metrics is that while the circle in 
\eqref{exacttdco} still shrinks to zero size at $t=0$, the circle in 
\eqref{dualexacttdco} that before was growing indefinitely, causing the 
curvature to diverge, now reaches a maximum size at $t=0$, of order 
$k/\sqrt{\alpha'}$. Note that T-duality of these two metrics is not 
realized simply as inversion of the size of the circle. This is as 
expected because $\alpha'$ corrections should not commute with small/large 
duality in string theory.

We interpret this as follows. The metric \eqref{exacttdco} governs
ordinary particles or pointlike strings moving on our two-dimensional
cosmology; such objects follow geodesics. The metric has a conical
singularity of Milne type near $t=0$, and is not geodesically complete. 
From the orbifold point of view, and also from the WZW perspective 
\cite{kr}, this suggests the presence of extra regions of spacetime with
closed timelike curves (called ``whiskers'' in ref.\ \cite{eliezer}).
On the other hand, the T-dual metric \eqref{dualexacttdco} governs 
pointlike strings in the T-dual space, which from the original point of 
view are winding strings. The non-singularity of this metric signals
that strings wound around the $r$-direction in our two-dimensional
cosmology \eqref{exacttdco} might propagate through without too much 
difficulty.

We thus see an inherently stringy mechanism that makes the origin of this
cosmological model look less singular than appears for ordinary particles,
and thereby constitutes a partial answer to the question raised above.
But clearly, the picture is far from being complete. We end this
section with pointing out a few more open issues.
\nxt
One might object that the metric \eqref{exacttdco} in the original picture
before T-duality still has a shrinking circle, and therefore we have not 
``resolved'' the singularity. However, this is as in ordinary Euclidean
orbifolds, even those that can actually be resolved to a smooth manifold,
such as the famous $A_1$ singularity which can be blown up to the
Eguchi-Hanson space. The only reasonable metric one can write down for the
orbifold is the singular one. String theory is only non-singular because
twisted/winding modes do not care so much about the original metric.
\nxt
One might ask the question what happens if we throw an ordinary
particle or a string without winding into this cosmology. In particular,
it was shown in \cite{hopo} that such a particle causes an insurmountable
amount of backreaction on the geometry. We do not know how this could
be avoided in our picture.
\nxt
It might be possible and interesting to see directly a change of the 
metric \eqref{dualmilne} for the T-dual of Milne for which the circle near
$t=0$ has a finite size. It would be also interesting see what other 
``exactly solvable'' cosmological backgrounds, such as the Nappi-Witten 
model \cite{nawi}, look like when treated in this way.
\nxt
Our discussion has been purely in the context of the bosonic string. In
particular, the beta functions that are solved to low order by the above
metrics are those for the bosonic $\sigma$-model. For the superstring,
worldsheet supersymmetry changes the structure of the beta functions,
and it is claimed that actually the tree-level metrics are exact, at
least for Euclidean spacetime signature. An obvious question then is whether 
this is still the case for the Minkowskian spacetime signature, and, if
the answer is positive, why adding the fermions has such a drastic effect 
on the way the winding modes probe the geometry.    
\nxt
An important issue that we have been suppressing so far concerns the
strength of the string coupling $g_s$ near the singularity, \ie, whether
a perturbative description can be expected at all. Clearly, the dilaton in
\eqref{exacttdco} is finite and constant near $t=0$, but blows up in
the T-dual version \eqref{dualexacttdco}, which includes $\alpha'$ corrections.
We could now perform S-duality $g_s\to 1/g_s$ in order to end up with a 
perturbative string description. However, the string frame metric of the 
target space becomes $ds_{\it string}^2\to 1/g_s\, ds_{\it string}^2$, so 
that now the full target space shrinks to zero size with a curvature 
singularity at $t=0$. Thus it appears that the our resolution of the locally 
Milne singularity results in a Big Crunch singularity instead\footnote{We 
would like to thank Gary Horowitz and Joe Polchinski for discussions on 
this.}.

\section{Summary}
\label{summary}

In this paper, we have presented a number of attempts at time dependence
in string theory and supergravity, with an emphasis on some practical
computations. Let us summarize our main results here.

Deformations of the gauge/gravity correspondence are a convenient tool
for studying various phenomena associated with singularities in
supergravity backgrounds. As we have shown here, following up on 
\cite{b0203}, these deformations also gives a very controllable way of 
introducing time dependence in supergravity and the dual field theories. 
For example, chiral symmetry restoration in de Sitter space can be easily
studied in the holographic dual. From a phenomenological point of view,
however, it is somewhat unfortunate that deformations other than de 
Sitter space become technically rather involved.

A more straightforward way of introducing time dependence is through
analytical continuation. Although one loses some control over stability
and well-definedness, it is by far the simplest tool to generate 
time-dependent solutions of supergravity. One of the practical
obstacles in applying the procedure in supergravity is the reality
of fluxes. In this paper, we have pointed out that there are in fact
some ways in which analytical continuation connects backgrounds with
real fluxes only. In principle, this allows the construction of a wide
variety of time-dependent backgrounds, of which we have presented a
few simpler ones here.

Analytical continuation is also a useful tool at the level of the
string worldsheet. In particular, exactly solvable string worldsheets
with Euclidean target space allow at least a formal continuation to
Minkowskian signature. Under the assumption that one can carry over
also certain non-perturbative worldsheet results, we have argued here
that they can teach us something about how strings perceive simple
cosmological singularities such as the Milne singularity.

\begin{acknowledgments}
We have greatly benefited from discussions with 
Justin David, Gary Horowitz, Shamit Kachru, Kirill Krasnov,
Jim Liu, Joe Polchinski, Radu Roiban, Christian R\"omelsberger, 
Eva Silverstein and  Arkady Tseytlin. The research of A.B.\ was supported 
in part by the NSF under Grant No. PHY00-98395 and PHY99-07949. The research 
of P.L.\ and J.W.\ was supported in part by the NSF under Grant No. 
PHY99-07949. P.L.\ would like to thank the KITP at Santa Barbara for 
hospitality while this work was being completed.
\end{acknowledgments}

\begin{appendix}

\section{Chiral symmetry restoration of MN model in de Sitter}
\label{appa}

In sections \ref{desitter} and \ref{mndesitter} we argued that the chiral 
symmetry that is broken in the nonsingular supersymmetric MN model 
\cite{mn0008} should be restored in a de Sitter background for sufficiently 
large Hubble parameter. In this appendix we present details supporting the 
claim.   

The appropriate equations describing de Sitter deformation of the 
MN model are given by \eqref{mngs} and \eqref{mncons2}, with (a) 
type boundary condition of \eqref{cases}. Unbroken chiral symmetry 
in the gauge theory implies $a\equiv 0$ in the dual supergravity. 
Solutions regular as $\r\to 0$ are characterized by three parameters 
$H$, $G_0$, and $g_0$, with asymptotics
\begin{equation}
\begin{split}
F^2&=n H^2 \r^2\left(1-\frac{16 G_0^2+8 G_0-1}{240 G_0^2}\r^2+O(\r^4)
\right)\cr
G^2&=G_0+\frac{4 G_0 -1}{20 G_0}\r^2+O(\r^4)\cr
g_s^{-2}&=g_0^{-2}\left(1+\frac{16 G_0^2+1}{40 G_0^2}\r^2+O(\r^4)\right)\,.
\end{split}
\eqlabel{chiral0}
\end{equation}
Here, $G_0^{1/2}$ is the size of the $S^2$ at $\r=0$, and $g_0$ is the string
coupling at $\r=0$. It seems that the important parameter in
\eqref{chiral0} is $G_0$, and not $H$, which we have claimed controls
chiral symmetry restoration. However, as we discuss below, the Hubble parameter 
$H$ which fixes the scale of the gauge theory background, is related to the 
first two parameters by the scale dependence of the gauge theory coupling. For 
now, we analyze the $G_0$ dependence of the solutions.

Starting from \eqref{chiral0}, we expect that as $\r\to \infty$ (UV 
from the gauge theory perspective), we recover the asymptotics of the 
flat LST for the physical solution representing the chirally symmetric 
phase of the MN model in the de Sitter. It turns out that the 
asymptotics at $\r\to\infty$ depend on whether $G_0<\frac{1}{4}$ or 
$G_0>\frac{1}{4}$. In the first case, the solution develops a singularity 
at some finite $\r=\r^*(G_0)$, while in the latter case we indeed find 
the flat LST asymptotics.

We begin with the special case $G_0=\frac{1}{4}$. With 
\eqref{chiral0} we find from \eqref{mngs} $G\equiv \frac{1}{2}$. 
Introducing a new radial coordinate
\begin{equation}
r\equiv \frac{F}{n^{1/2} H} \,,
\eqlabel{newmn}
\end{equation}
we find from \eqref{mngs} the following first order equation for 
$f\equiv r [\ln g_s]'$
\begin{equation}
f'=-\frac{(f-1)(f-3)(3f+4 r^2)}{r (2 r^2+3)}\,.
\eqlabel{phimn}
\end{equation}
The boundary 
condition \eqref{chiral0} then translates into 
\begin{equation}
f=-\frac{4}{5} r^2-\frac{16}{175}r^4+O(r^6)\,.
\eqlabel{boundf}
\end{equation}
Though we can not find an exact analytical solution to \eqref{phimn},
the $r\to \infty$ asymptotic can be easily extracted
\begin{equation}
f(r\to \infty)=-\frac{4}{3}r^2+1+O(1/r^2) \,.
\eqlabel{dilass}
\end{equation} 
The complete string frame metric of the special $G_0=\frac{1}{4}$ 
solution reads 
\begin{equation}
\begin{split}
ds_{st}^2&=n H^2 r^2\left[-dt^2+\frac{1}{H^2}\cosh^2 H t\ d\O_3^2\right]
+n \left(\triangle^2 dr^2+\frac{1}{4} ds_{\tilde{T}^{1,1}}^2
\right)\cr
&\equiv n\left(r^2 dS_4^2+\triangle^2 dr^2\right)+
\frac{n}{4}\ ds_{\tilde{T}^{1,1}}^2 \,,
\end{split}
\eqlabel{fullmetspe} 
\end{equation}
where 
\begin{equation}
\triangle^2= \frac{(f-1)(f-3)}{3+2  r^2} \,,
\eqlabel{trm}
\end{equation}
and 
\begin{equation}
ds_{\tilde{T}^{1,1}}^2\equiv d\ttheta^2+\sin^2\ttheta d\tphi^2
+d\theta^2+\sin^2\theta d\psi^2 + (d\phi+\cos\theta d\psi-\cos\ttheta 
d\tphi)^2 \,.
\eqlabel{t11}
\end{equation}

The $\tilde{T}^{1,1}$ metric above along with the corresponding
NS-NS background  \eqref{Hmn} was identified in \cite{zt0007} as 
target space of the simplest representative in the class of 
$\frac{\SU(2)\times \SU(2)}{\U(1)}$ coset sigma models introduced in 
\cite{mncosets}. In \cite{zt0007} conformal invariance of this coset 
was checked in the 3-loop approximation, and the expectation is that 
this background is an exact NS-NS string solution to all orders in 
$\a'$. The same coset appears in the ``special Abelian solution''
of the  non-BPS excitation of the MN model \cite{gtv0108}.  
This ``special Abelian solution'' has an infrared singularity due to the 
linear dilaton. Thus we are led to the conclusion that our 
special $G_0=\frac{1}{4}$ solution is a de Sitter regularization 
of the singular ``special Abelian solution'' of  \cite{gtv0108}.
The exact central charge for the $\tilde{T}^{1,1}$ coset is 
\cite{gtv0108}
\begin{equation}
c_{\tilde{T}^{1,1}}=5\cdot \frac 32-\frac{12}{n} \,.
\eqlabel{cc}
\end{equation} 
Thus, the sigma-model corresponding to the time-dependent target 
space metric
\begin{equation}
ds_{5}=n\left(r^2 dS_4^2+\triangle^2 dr^2\right)
\eqlabel{5target}
\end{equation}
and dilaton 
\begin{equation}
\Phi=\ln g_0+\int_0^r d\xi\ \frac{f(\xi)}{\xi}
\eqlabel{dilresult}
\end{equation}
with $f$ being a solution of \eqref{phimn} with boundary condition 
\eqref{boundf}, must have central charge
\begin{equation}
c_5=5\cdot \frac 32+\frac{12}{n} \,.
\eqlabel{cc5}
\end{equation} 
The latter can be easily verified as asymptotically, 
\eqref{5target}, \eqref{dilresult} is a linear dilaton 
background.  

It would be very interesting to compute $\a'$ corrections to \eqref{5target}
and to identify the appropriate CFT. The ``special Abelian solution''
of \cite{gtv0108} was shown to be unstable. It would be interesting to 
check whether its de Sitter regularization, realized by our special 
$G_0=\frac{1}{4}$ solution, is stable. 

We now proceed with the general $G_0\ne \frac{1}{4}$ case. We could 
not find an exact analytical solution for general $G_0$. For numerical 
analysis we find it convenient to use a new radial coordinate $r\equiv 
\r^2,\ r\in [0,+\infty)$, and introduce $f$, $g$, and $h$ by
\begin{equation}
\begin{split}
F^2&\equiv n H^2 f\cr
G^2&\equiv g\cr
g_s^{-2}&\equiv g_0^{-2} h \,.
\end{split}
\eqlabel{change}
\end{equation} 
With the boundary conditions \eqref{chiral0} the only parameter in
the numerical integration becomes $G_0$. We find qualitatively 
different $r\to\infty$ asymptotics depending on the value for $G_0$. 
Namely, for $0<G_0<\frac{1}{4}$, we have,
\begin{equation}
\begin{split}
&h(r\to r^*)\to +\infty\cr
&g(r\to r^*)\to 0\cr
&f(r\to r^*)\to {\rm const} \,,
\end{split}
\eqlabel{smg0}
\end{equation}
where $r^*\equiv r^*(G_0)$ is finite, and for $G_0>\frac 14$, we find
\begin{equation}
\begin{split}
&h(r\to \infty)\to +\infty\cr
&g(r\to \infty)\to +\infty\cr
&f(r\to \infty)\to +\infty \,.
\end{split}
\eqlabel{lgg0}
\end{equation}

We can now check analytically whether there are asymptotic solutions of
\eqref{mngs} which have the same qualitative behavior as predicted by
the numerical analysis \eqref{smg0} and \eqref{lgg0}. Indeed, solutions 
with such leading asymptotics exist. Explicitly, we find first
\begin{equation}
\begin{split}
h&=\frac{(4\delta_1^2 r^{*})^{1/4}}{\sqrt{r^*-r}}\left[1+O(r^*-r)\right]\cr
g&=\frac{1}{2}\sqrt{1-\frac{r}{r^*}}\left[1+O(r^*-r)\right]\cr
f&=\delta_2\left[1+O(r^*-r)\right] \,.
\end{split}
\eqlabel{ansm}
\end{equation}
If we identify this solution as corresponding to the $0<G_0<\frac 14$ 
boundary condition, then $\delta_1,\delta_2,r^*$ will be dependent on 
$G_0$. Also, there is a solution
\begin{equation}
\begin{split}
h&=\delta_1 e^{2 r^{1/2}}\left[1+o(1)\right]\cr
g&\to  r^{1/2}\left[1+o(1)\right]\cr
f&\to 3 r^{1/2}\left[1+o(1)\right] \,,
\end{split}
\eqlabel{anlg}
\end{equation}
which we identify as corresponding to the $G_0>\frac{1}{4}$ boundary 
condition, with $\delta_1$ depending on $G_0$.

To summarize, we have shown using numerics and asymptotic analysis 
that a chirally symmetric phase (nonsingular solution to \eqref{mngs} 
with \eqref{chiral0} boundary condition) exists only for
\begin{equation}
G_0>G_{\rm critical}\equiv \frac{1}{4} \,.
\eqlabel{critg0}
\end{equation}
Although this is not an analytical proof, it strongly support the
picture of a chiral symmetry restoration phase transition in the de
Sitter deformed MN model. So far, this phase transition appears to
occur as the (dimensionless) parameter $G_0$ is varied. We now argue 
that in fact, the relevant physical parameter is the Hubble scale $H$.

In addition to $G_0$ satisfying \eqref{critg0}, nonsingular deformations
are characterized, naively, by two more parameters $g_0$ and $H$.
This is one parameter too many: from the perspective of LST wrapped on
$S^2$, we expect only two continuous parameters: $g_0$ (the dilaton
parameter in the supersymmetric MN model) and the deformation parameter,
$H$. In the remainder of the appendix we argue that for fixed $\{g_0,H\}$,
$G_0$ is not actually an independent parameter, but rather
\begin{equation}
\ln H\sim G_0^2 \,,
\eqlabel{GHrel}
\end{equation}
and thus, the non-singularity condition \eqref{critg0} translates into 
a lower bound on $H$. 

One may observe that rescaling $t\to\tau\equiv H t$ entirely removes the 
$H$ dependence from the background. But this is not very illuminating, as 
in doing so we are changing the reference energy scale from the LST 
perspective. Instead, we extract the relation \eqref{GHrel} from the gauge 
theory beta-function, keeping the effective four-dimensional gauge theory 
strong coupling scale (and the string scale) fixed. We don't know the 
beta-function of gauge theories in de Sitter space,  but for asymptotically 
free gauge theories the perturbative beta-function should be roughly
the same as the standard Minkowski one. The reason is that it is related 
to the short distance dynamics of the theory, and we expect de Sitter 
deformation to be irrelevant at small scales. We will use the Minkowski 
beta-function with the understanding that at best only qualitative physics 
would come out correct. Now recall that the open string interpretation of 
the MN model is LST compactified on $S^2$. At low energies the effective 
description is in terms of $\caln=1$ SYM theory with gauge coupling 
$g_4$ \cite{mn0008},
\begin{equation}
\frac{1}{g_4^2}=\frac{{\it Vol}_{S^2}}{g_6^2}=\frac {n G^2}{2\pi^2} \,,
\eqlabel{gaugecoup}
\end{equation} 
where $g_6$ is the D5-brane gauge theory coupling. In the supergravity 
dual, $G$ depends on $\r$ which is interpreted as the RG running of the 
effective four-dimensional coupling. Verifying \eqref{gaugecoup} requires 
understanding of UV/IR relation in the MN model, or, in other words, 
the translation of the radial direction $\r$ into the RG gauge theory 
scale $\mu$. This relation can not be defined unambiguously. Maldacena and 
Nu\~nez \cite{mn0008} used $\mu\sim e^{\r/2}$, and showed that
\eqref{gaugecoup} agrees with the gauge theory result $1/(g_4^2 n)\sim 
\ln\mu/\Lambda_{QCD}$ up to a numerical coefficient. In the MN model 
with unbroken chiral symmetry $G^2$ was singular at $\r=0$. This 
singularity reflects the Landau pole in the perturbative gauge theory 
coupling. On physical grounds we expect that the effective gauge theory 
Landau pole would be regulated in the de Sitter for large enough
Hubble parameter simply because the RG running of the gauge coupling 
stops at $\mu=\mu_{\it IR\;cutoff}\sim H$. This fits nicely with the
supergravity result for the de Sitter deformed MN model where $G^2\ge 
G_0^2>0$. Provided we can use \eqref{gaugecoup} when $H\ne 0$, we are 
led to \eqref{GHrel}.

Needless to say, our heuristic arguments can not replace a precise 
computation. However, such a computation requires (at the very 
least) an understanding of the precise UV/IR relation\footnote{The
UV/IR connection is very subtle even for the gauge/gravity correspondence 
for D5/NS5 branes \cite{pp9809}.} in the MN model and a precise 
understanding of perturbative (non-abelian) gauge theory dynamics in 
de Sitter space. Both problems are very difficult.  For 
recent progress on the first of them, see, \eg, \cite{bigazzi,merlatti}.
 
\end{appendix}

\end{document}